\documentclass[11pt]{article}

\usepackage{graphicx}
\usepackage{dcolumn}% Align table columns on decimal point
\usepackage{bm}% bold math
\usepackage[all]{xy}

\begin{document}

%\preprint{APS/123-QED}

\title{One particle self correlations for a separable representation
of the singlet spin state beyond standard Quantum Mechanics}

\author{Carlos L\'opez\\
Department of Mathematics and Physics, UAH\\
28873 Alcal\'a de Henares (Madrid) SPAIN\\
carlos.lopez@uah.es}

\date{\today}

\maketitle

%%%%%%%%%%%%%%%%%%%%%%%%%%%%%%%%%%%%%%%%%%%%%%%%%%%%%%%%%%%%%%%%%%%%%%%%%%%%%
\centerline{Abstract}

A new pure quantum state, isotropic in spin variables, is defined in an extended spin phase space beyond standard quantum mechanics. It allows to represent the entangled singlet state in separable form.  The statistical correlations between  Alice and Bob measurements become self correlations between hidden spin values for each particle, together with perfect anti correlation between spin values on the pair. Alice determines through measurement on her particle the value of spin  in some direction. Spin in another direction is inferred from Bob measurement on the companion. Bell's inequalities are violated because of the wave like behaviour of quantum systems. In full analogy with the two slit experiment, interference terms between spin field components appear  determining the contextual character of quantum distributions of probability.

\vskip 0.7 truecm


\vskip 0.4 truecm
                                

%%%%%%%%%%%%%%%%%%%%%%%%%%%%%%%%%%%%%%%%%%%%%%%%%%%%%%%%%%%%%%%%%%%%%%%%%%%

\newpage

\section{\label{sec1}After Bell}

Violations of Bell's type inequalities in spatially se\-pa\-rated measurements\cite{Aspect} have been empirically tested beyond any reasonable doubt\cite{tests,t3,t4,t5}. All relevant loopholes have been satisfactorily closed, and the predicted quantum correlations  confirmed. It is time to look for physical evidence of the non local influence between measurement events. However, known interactions are mediated by physical systems, either particles following time--like or light--like paths  or  distributed fields evolving relativistically.
Space--like curves as paths of particles are discarded because they  have a frame dependent time orien\-ta\-tion, according to relativity.
The value of a field at a space time event depends on its values along a spatial sheet inside the past light cone; in other words, it commutes with values of the field at spatially separated events. Even in case of time--like separation between measurements it is unlikely the existence of a media\-ting system connecting them  without observable decay for increasing distance and with  other systems in between that do not shield its propagation.   

We can, alternatively,  go beyond the standard formulation of Quantum Mechanics (SQM) and  develop an explicitly local, separable description of entangled states for composites. 
The celebrated EPR paper\cite{EPR} about incompleteness of SQM,  the quantum potential in Bohm mechanics\cite{Bohm}, the  
analysis of Renninger\cite{ren} of the wave particle duality in an interferometer,  inconsistency between SQM and the action reaction principle\cite{ARP,ARP2}, among other considerations\cite{other,o2,o3}, are enough arguments to explore the possibility of a formulation of  Quantum Mechanics in extended phase spaces\cite{EQM}(EQM). 

States of quantum systems could be described by some $[x,\Phi]$, $x$ commuting and non commuting variables of a corpuscular subsystem and 
$\Phi$ an accompanying de Broglie\cite{dBro} (or pre--quantum,  
sub quantum\cite{subq,s3}) 
field. In the double field $\Phi$--$\Psi$ model, the distribution of amplitude 
$\Psi = R exp(i\theta)$ is a statistical representation of an ensemble of composite systems, $R^2$  distribution of probability for variables $x$ of the particle and 
$\bigtriangleup \theta$ relative phases between field components.

No go theorems\cite{nogo,n2,n3,n4,n5} are dead end paths for the pursuit of an extended phase space in EQM. The existence of global, non contextual distributions of probability $P(x)$ in spaces of (so called) hidden variables for the particle and whose marginals match the quantum distributions are mathematically forbidden. But these  distributions would ignore the accompanying field and its interaction with the particle.   Obviously, models that do not fulfil some hypothesis of these no go theorems are not ruled out\cite{EQM}. 
The empirical fact is that entanglement appears exclusively after  local interaction in the past between the correlated systems, or it is conditional to some intermediate interaction if both systems of interest have not been in contact in the past.

In this letter, a local description for the quantum co\-rre\-la\-tions in the singlet spin state is formulated in the framework of an extended phase space  for spin variables. In section II, quantum distributions of amplitude,  not classical distributions of probability,  describe all   standard pure quantum states, each one eigenstate of some spin ope\-ra\-tor, and  a new  pure quantum state isotropic in spin, which has no counterpart in SQM. It is this ingre\-dient of the formalism, the use,  as in SQM, of quantum amplitudes in the extended space instead of classical proba\-bi\-li\-ties, which allows to overcome the thesis of no go theorems. The calculation of marginal amplitudes through projection over the standard phase spaces, followed by Born rule, reproduces the results of SQM, that is, the correlations between a previously known (the eigenvalue) and a measured value of spin. Correlations between hidden va\-lues of spin in two arbitrary  directions can be consistently computed in the isotropic state. In the singlet state of a composite, section III, two values of spin can be determined for each particle. One is obtained through direct measurement and the other is inferred from the perfect anti correlation with the measured companion. The state of the composite is separable, each particle is in the new isotropic spin state, with its associated individual self correlations. This formalism can be  relevant for the  ontological interpretation of Quantum Mechanics, the ensemble character of pure quantum states.

\section{\label{sec2}The isotropic spin state}

If locality is assumed for joint and spatially separated measurements on the singlet spin state, perfect anti correlation between outputs for any common, freely and independently chosen by Alice and Bob, direction ${\bf n}$ of measurement implies that a complete representation of the physical state is characterised by all values of spin.  If,
to avoid mathematical complications,  we consider  a finite set of directions
$\{{\bf n}_1,{\bf n}_2,\ldots,{\bf n}_N \}$,   values
$(s_1^{\alpha},s_2^{\alpha},\ldots,s_N^{\alpha})_{\nu}$ and 
$(s_1^{\beta},s_2^{\beta},\ldots,s_N^{\beta})_{\nu}$ for each pair of jointly generated  particles $(\alpha_{(\nu)},\beta_{(\nu)})$,
$\nu \in \{1,2,\ldots\}$, are fixed from the generation event, fulfilling 
$s_{j(\nu)}^{\alpha} +$  $s_{j(\nu)}^{\beta} = 0$.
Three independent values as $s_x$, $s_y$, $s_z$
do not determine the other variables of spin, e.g. $s_{\theta}$
for the magnitude (operator) $S_{\theta} = \cos(\theta) S_x + \sin(\theta) S_y$.
The functional relations between non commuting operators are not fulfilled by 
their eigenvalues, $s_{\theta} \neq \cos(\theta) s_x + \sin(\theta) s_y$. In SQM, the dimension of the phase space is lower than in Classical Hamiltonian Mechanics,  e.g. position and momentum variables $\{(q,p)\}$ are restricted to $\{q\}$ (resp. $\{p\}$) in the position (momentum) representation. The phase space of EQM has higher dimensions than its classical counterpart, according to the infinite degrees of freedom of the accompanying field.

Let us consider the extended spin phase space 
${\cal P}h=\{(s_1,\ldots,s_N)|s_k=\pm\}$, $|{\cal P}h|=2^N$, associated to an elementary spin $1/2$ particle. Bell's inequalities state that for $N>2$ there are not global, non contextual distributions of probability 
on ${\cal P}h$, describing a classical statistical ensemble from which the quantum probabilities for the singlet could be obtained.

\[
P_{QM}(s_1^{\alpha},s_2^{\beta})=\frac {1}{4}(1-s_1^{\alpha}s_2^{\beta}
{\bf n}_1\cdot{\bf n}_2) =
\] 

\[
= \frac {1}{4}(1+s_1^{\alpha}s_2^{\alpha}
{\bf n}_1\cdot{\bf n}_2) = P_{EQM}(s_1^{\alpha},s_2^{\alpha})
\]
can not be reproduced by a global distribution of pro\-ba\-bility 
$P_{Cl}(s^{\alpha}_1,$ $\ldots,s^{\alpha}_N)$ through marginals
$\sum _{l\neq1,2}\sum _{s_l}$ $P_{Cl}(s^{\alpha}_1,$ $\ldots,s^{\alpha}_N)$.
The existence of a classical probabilistic mixture $P_{Cl}$ of physical states with hidden variables, representing an ensemble quantum state, is a ``natural'' hypothesis
systematically considered  in the literature of no go theorems. However, it is not 
unavoidable, and interference phenomena as in the paradigmatic two slit experiment
point to the need of other mathematical tools.
An alternative algorithm must be applied in
${\cal P}h$, able to reproduce the quantum distributions for a
statistical sample of measurements over the same pure/ensemble quantum state.
Let us apply the  ``quantum way'', a distribution of amplitude of probability 
$Z(s_1,\ldots,s_N)$, $Z: {\cal P}h \to K$ (in the spin phase space, $K$ will be 
the set of ima\-gi\-na\-ry quaternions). We can mimic the paradigmatic two slit experiment
and obtain marginals for the distribution of amplitude $Z$

\[
Z(s_j) =  \sum _{l\neq j}\left( \sum _{s_l} Z(s_1,\ldots,s_N) \right)
\]
Applying now Born rule, we get the probabilities

\[
P(s_j) = \frac {|Z(s_j)|^2}{|Z(+_j)|^2+|Z(-_j)|^2} \, ,
\]
where there will appear generically interference terms in the 
squared sum of amplitudes. Compare it  with

\[
\Psi(x_0,y_0) = \Psi _L(x_0,y_0) + \Psi_R(x_0,y_0)
\]

\[
P(x_0,y_0) = \frac {|\Psi(x_0,y_0)|^2}{\sum _{(x,y)}|\Psi(x,y)|^2} \, ,
\]
where $(x,y)$ are the position variables at the final screen of the two slit experiment, 
and $\Psi(x_0,y_0) = \Psi _L(x_0,y_0) + \Psi_R(x_0,y_0)$ is the marginal amplitude, 
sum of  left and right slit field  components. Interference terms in $|\Psi(x_0,y_0)|^2$ are here responsible of the diffraction pattern. Similarly, interference terms in 
$|Z(s_j)|^2$ are responsible of the contextual character of quantum distributions of probability, i.e., its dependence on the field components allowed by the physical context.

Formal distributions of probability for correlated va\-lues of spin in two arbitrary
directions ${\bf n}_j$ and ${\bf n}_k$ can be similarly determined, although in different, alternative ways.
One of these values remains necessarily counterfactual because of the incompatibility of joint measurements; measurement of $s_j$ unavoidably perturbs the previous value of  $s_k$. From the marginals

\[
Z(s_j,s_k) =  \sum _{l\neq j,k}\left( \sum _{s_l} Z(s_1,\ldots,s_N) \right)
\]
we could formally define the joint, unobservable distribution

\[
\Pi (s_j,s_k) = \frac {|Z(s_j,s_k)|^2}
{\sum _{s'_j,s'_k}|Z(s'_j,s'_k)|^2} \, ;
\]
and the same definition can be generalized to $\Pi(s_j,$ $s_k,s_l)$, etc.
Generically, $P(s_j)$ $\neq$ $\Pi(s_j,+_k)$ $+$ $\Pi(s_j,-_k)$ 
because of the interference terms when Born rule is applied to a sum of 
amplitudes 

\[
|Z(s_j,+_k)+Z(s_j,-_k)|^2=|Z(s_j,+_k)|^2+|Z(s_j,-_k)|^2 +
\]

\[
+ \Big( Z^*(s_j,+_k)Z(s_j,-_k)+Z^*(s_j,-_k)Z(s_j,+_k)b \Big) _{\rm interf}
\]
We can interpret $P(s_j)$ and $\Pi(s_j,s_k)$ as corresponding to incompatible physical contexts, as in the two slit experiment.
Alternatively, we can also define conditional probabilities

\[
\Pi(s_k|s_j) = \frac {|Z(s_j,s_k)|^2}
{\sum _{s'_k}|Z(s_j,s'_k)|^2}
\]
from which

\[
\Pi (s_j;s_k) = P(s_j) \Pi(s_k|s_j)
\] 
and similarly for $\Pi (s_k;s_j)$.
Now, $P(s_j)=$ $\Pi(s_j;+_k)$ $+$ $\Pi(s_j;-_k)$ but generically
$\Pi (s_j;s_k)$ $\neq$ $\Pi (s_j,s_k)$ $\neq$ $\Pi (s_k;s_j)$. 
When values $s_j$ and $s_k$ are jointly observable, as in the singlet state, it must happen   that  $\Pi(s_j;s_k)=$  
$\Pi (s_j,s_k)=$ $\Pi(s_k;s_j)$, matching the observed $P(s_j,s_k)$. This will happen if the physical contexts associated to $P(s_j)$ and $P(s_j,s_k)$ are compatible.

Let us consider the quaternion 

\[
{\bf N}[{\bf n}]= ({\bf n}\cdot{\bf i}){\bf I} + 
({\bf n}\cdot{\bf j}){\bf J} + ({\bf n}\cdot{\bf k}){\bf K} \, ,
\]
with null real part, associated to a unit vector $\bf n$. Each spin state 
$(s_1,\ldots,s_N)$ will have a fixed as\-so\-cia\-ted  amplitude $Z$, sum of elementary amplitudes $s_j{\bf N}_j\equiv$
$s_j{\bf N}[{\bf n}_j]$, in analogy with the elementary amplitudes 
$e^{iS[{\rm path}]/\hbar}$ for virtual paths in the path integral formalism, 
$Z(s_1,\ldots,s_N) \equiv \sum _j s_j {\bf N}_j$
The physical context determines which {\it virtual spin} states  are considered, in the same way that different phy\-si\-cal configurations determine the virtual paths to be taken into account, e.g. in the two slit experiment. The SQM state  $|+_1>$, spin up in direction
${\bf n}_1$, can be prepared using a Stern--Gerlach apparatus
that splits the incoming trajectory into up and down spin output paths. The up path
does not have down spin field components, so that in the extended formalism
the ensemble state $|+_1\!>$ has associated distribution of amplitude
$Z_{+_1}(s_1,s_2,\ldots,s_N)$ where  $Z_{+_1}(-_1,\ldots)\equiv 0$.
The marginals become $Z_{+_1}(-_1)=0$, $Z_{+_1}(+_1) = 2^{N-1}{\bf N}_1$,
$Z_{+_1}(s_2) = 2^{N-2}({\bf N}_1+s_2{\bf N}_2)$. When $s_2$ is measured
$\pm_j$ terms interfere for $j\geq 3$.
These marginal amplitudes determine the associated observable probabilities, $P_{+_1}(-_1)=0$, $P_{+_1}(+_1)=1$, as well as $P_{+_1}(s_2)=(1+s_2{\bf n}_1\cdot{\bf n}_2)/2$, where
the relations

\[
{\bf N}^*=-{\bf N} \quad {\bf N}^2=-1 \quad {\bf N}^*_1{\bf N}_2=
{\bf n}_1\cdot{\bf n}_2 - {\bf N}[{\bf n}_1\times {\bf n}_2]
\]
have been used. The SQM distributions are reproduced.

When the context allows both spin up and down field components in all directions the corresponding state $S_0$ becomes isotropic, with distribution of amplitude $Z_0$ containing all components in ${\cal P}h$, $Z_0\equiv Z$, and distributions of probability $P_0(s_j)=1/2$ for all $j$. This quantum state has no counterpart in the Hilbert space of SQM, where every vector of state is up eigenstate for the spin operator in some direction. A classical mixture like

\[
\rho=\frac {1}{2} |+_1><+_1| + \frac {1}{2} |-_1><-_1|
\]
reproduces the isotropic distribution too, but it
has different ontological content; $\rho$ represents two sub--ensembles of pure states, 
$|-_1>$ and $|+_1>$, each one lacking the other field components, while $S_0$ contains all of them which can interfere. $S_0$ and $\rho$ are associated to different physical contexts.

Formal distributions of probability for two or more va\-lues of spin are obtained  through marginal amplitudes and Born rule,

\[
\Pi_0(s_1,s_2) = (1+s_1s_2{\bf n}_1\cdot{\bf n}_2)/4 \, ,
\] 
proportional to the (not normalized) squared marginal amplitude

\[
| 2^{N-2} \left( s_1 {\bf N}_1 + s_2 {\bf N}_2 \right) |^2
\]
as well as $\Pi_0(s_1,s_2,s_3) = $

\[
\frac {1}{24}(3 + 2s_1s_2 {\bf n}_1\cdot{\bf n}_2
+ 2s_1s_3 {\bf n}_1\cdot{\bf n}_3 + 2s_2s_3 {\bf n}_2\cdot{\bf n}_3)
\]
proportional to 

\[
|2^{N-3} \left( s_1 {\bf N}_1 + s_2 {\bf N}_2 + s_3 {\bf N}_3 \right) |^2
\]
$\Pi _0(s_1,s_2)$ is not observable, but it is consistently defined: $\Pi_0(s_1,+_2)+$ 
$\Pi_0(s_1,-_2)$ $=P_0(s_1)$,
and $\Pi_0(s_1;s_2)$ $=\Pi_0(s_1,s_2)$, so that we can consider a ``classical'' distribution $P_0(s_1,s_2)$. On the other hand, a 
$P(s_1,s_2,s_3)$ is not consistently defined,

\[
\Pi_0(s_1,s_2,+_3) + \Pi_0(s_1,s_2,-_3) \neq P_0(s_1,s_2)
\]
Notice the analogy with the two slit experiment

\[
P(x,y,L) + P(x,y,R) \neq P(x,y)
\]

\section{\label{sec3}The singlet state}

Two particles $\alpha$ and $\beta$ are jointly generated in the singlet spin state

\[
|S_{\rm singlet}> = 
|+_1>^{\alpha}\otimes |-_1>^{\beta}
- |-_1>^{\alpha}\otimes |+_1>^{\beta} 
\]
Each particle (marginal) density is isotropic in spin va\-ria\-bles, $P(s_j^{\alpha})=$ $1/2$ for all directions $j$; no individual pure quantum  state of $\alpha$ in the two dimensional Hilbert spin space of SQM can represent it. In the usual interpretation of SQM, with one to one correspondence between physical and pure quantum states, a separable description of the singlet is not possible. On the other hand, in the extended phase space where pure quantum states, distributions of quaternion amplitudes, represent ensembles of physical states, the $\alpha$-$\beta$ correlation applies to  jointly generated pairs 
$(\alpha_{(\nu)},\beta_{(\nu)})$ and not to isotropic spin states $S_0^{\alpha}$ and 
$S_0^{\beta}$, which describe statistical ensembles for each particle separately.
It is obvious that there is not correlation between pairs of outputs for Alice and Bob measurements over particles $\alpha_{(\nu)}$ and $\beta_{(\nu')}$ belonging to different pairs
$\nu\neq\nu'$.  Correlations apply to jointly generated particles $s_{j(\nu)}^{\alpha}+s_{j(\nu)}^{\beta}=0$.
The singlet state is expressed in separable form $S_{\rm singlet} =$
$S_0^{\alpha}\otimes_{\rm corr}S_0^{\beta}$ if
$\otimes_{\rm corr}$ is understood as the perfect anti correlation between 
jointly generated pairs. Each particle, if we ignore the companion, is  in the pure state $S_0$ of EQM. 

Equivalently, a distribution $Z_{\rm singlet}$ can be defined on the  subset 
${\cal P}h_{\rm corr}\subset$ ${\cal P}h_{\alpha}\times$ ${\cal P}h_{\beta}$ defined by the correlation equations, or

\[
Z_{\rm singlet}\Big(
(s_1^{\alpha},\ldots,s_N^{\alpha}),
(s_1^{\beta},\ldots,s_N^{\beta})\Big)\equiv 0
\]
outside ${\cal P}h_{\rm corr}$ ($s_{j}^{\alpha}+s_{j}^{\beta}\neq0$ for some $j$) and

\[
Z_{\rm singlet}\Big(
(s_1^{\alpha},\ldots,s_N^{\alpha}),
(-s_1^{\alpha},\ldots,-s_N^{\alpha})\Big)\equiv 
\]

\[
\equiv  Z_0(s_1^{\alpha},\ldots,s_N^{\alpha}) = - Z_0(s_1^{\beta},\ldots,s_N^{\beta})
\]
The $\alpha$ ($\beta$) marginal of $Z_{\rm singlet}$, when projecting 
from  ${\cal P}h_{\rm corr}$ onto ${\cal P}h_{\alpha}$ (${\cal P}h_{\beta}$), becomes trivially 
(there is only one non vanishing term in the fibre of the projection) the isotropic 
$S_0^{\alpha}$ ($S_0^{\beta}$), i.e., they are pure quantum states and not mixtures as
the marginals of  the density $\rho_{\rm singlet}=$ $|S_{\rm singlet}><S_{\rm singlet}|$.
The formal distribution $\Pi _0 (s^{\alpha}_{j(\nu)},s^{\alpha}_{k(\nu)})$  is observable.
Recall it is consistent with $\Pi _0 (s^{\alpha}_{j(\nu)};s^{\alpha}_{k(\nu)})$ and 
$\Pi _0 (s^{\alpha}_{k(\nu)};s^{\alpha}_{j(\nu)})$, defining a classical distribution. The second value of spin is inferred from the output of Bob measurement over $\beta _{(\nu)}$, without perturbing the state of $\alpha _{(\nu)}$. It means we observe (infer) the 
 value of $s^{\alpha}_{k(\nu)}$ previous to  measurement of $s^{\alpha}_{j(\nu)}$.
Phy\-si\-cal splitting into $\pm_k$ spin field components of 
$\beta _{(\nu)}$, at Bob's apparatus, does not perturb Alice's $\alpha _{(\nu)}$ particle. Splitting of  the $\alpha _{(\nu)}$ $\pm_j$ spin field components maintains on each branch both $\pm_k$ (and  other $\pm_l$) spin components of the 
total spin field, which  interfere.
Both ${\bf n}_j$ and ${\bf n}_k$ are freely and independently chosen by Alice and Bob. The correlations in each individual isotropic spin state $S_0$ are an inner property of each particle  separately. The predicted distributions of probability obtained from $S_0$
through marginal amplitudes and Born rule, observable because of perfect anti correlation,
match the SQM predictions for the  entangled singlet state.

When considering a third direction, interference in

\[
|Z_0(s_1,s_2,+_3)+Z_0(s_1,s_2,-_3)|^2
\]
does not vanish. A global classical distribution 
of probability $P_{Cl}(s_1,\ldots)$ does not exist, according to Bell's inequalities. As in the two slit experiment,
field components of hidden, not measured magnitudes are superposed and interfere.
The only relevant distinction between both physical processes is that $x$ and $y$ position coordinates at the final screen commute and can be jointly measured on an individual particle, while $s_1$ and $s_2$ do not commute and one of them can only be inferred from measurement on the correlated companion. Counterfactual values are widespread in Physics, and our degree of confidence in them is linked to our confidence (empirically grounded) in the applied theory, in this case Quantum Mechanics. The property of consistency depends on the quantum state, here the isotropic spin $S_0$. We could calculate in an orthodox $Z_{+_1}$ state
formal joint or conditional correlations between $s_2$ and $s_3$, but $\Pi_{+_1}(s_2;s_3)$
$\neq \Pi_{+_1}(s_2,s_3)$ $\neq \Pi_{+_1}(s_3;s_2)$ are incompatible. $s_2$ and $s_3$ variables are not jointly observable.

The contextual character of the quantum distributions is already present in the paradigmatic two slit ex\-pe\-ri\-ment. Two non vanishing probabilities $P_R(x,y)$ and $P_L(x,y)$, applied each to the physical context with one slit open and the other closed, do not add to the distribution $P(x,y)$ in the third context with both slits open. Wave superposition and interference, a typical phenomenon for distributed fields, is behind this contextual behaviour of the quantum probabilities, and suggest to interpret
elementary particles  as composites
made of a corpuscular system and a distributed, relativistically (locally, causally) evolving field.
The same phenomenon, applied to spin field components, is found in the isotropic spin state,
which is a pure quantum state in EQM without counterpart in SQM.
Other entangled composites in SQM could also find 
a local, separable and  contextual representation through new states in 
adequate extended phase spacse of EQM.

\section{\label{ack}Acknowledgements}

Financial support from project MTM2015-64166-C2-1-P (Spain) is acknowledged.


\begin{thebibliography}{}


\bibitem{Aspect} A. Aspect, P. Grangier and G. Roger : 
``Experimental Test of Bell's Inequalities Using Time--Varying Analysers'',
PRL {\bf 49} (25) 1804--1807  (1982)

\bibitem{tests}  
J.-W. Pan {\it et al} : 
``Experimental test of quantum non\-lo\-ca\-li\-ty in three-photon Greenberger--Horne-–Zeilinger entanglement'',
{\em Nature} {\bf 403}, 515-519 (2000)


\bibitem{t3} B. Hensen {\it et al} : ``Loophole-free Bell inequality violation using electron spins separated by 1.3 kilometres'', {\em Nature} {\bf 526}: 682–686 (2015).


\bibitem{t4} Giustina, M. {\it et al} : ``Significant-Loophole-Free Test of Bell’s Theorem with Entangled Photons'', PRL {\bf 115} 250401 (2015).


\bibitem{t5} L. K. Shalm {\it et al} : ``Strong Loophole-Free Test of Local Realism''
PRL {\bf 115}, 250402 (2015).


\bibitem{EPR} A. Einstein, B. Podolsky and N. Rosen :
``Can quantum--mechanical description of physical reality be considered complete?'',
{\em Phys. Rev.} {\bf 47}, 777--780 (1935)

\bibitem{Bohm} D. Bohm : 
``A Suggested Interpretation of the Quantum Theory in Terms of `Hidden` Variables. I and II'', {\em Phys. Rev.} {\bf 85} 166-192 (1952)

\bibitem{ren} M. Renninger : ``Zum Wellen Korpuskel Dualismus'',
{\em Zeitschrift f\"ur Physik},
{\bf 136}, 251--261 (1953)

\bibitem{ARP} C. L\'opez : ``A local interpretation of Quantum Me\-cha\-nics'',
{\em Found Phys},  46: 484--504 (2016). 

\bibitem{ARP2} C. L\'opez : ``Relativistic locality and the action reaction principle predict de Broglie fields'', arXiv: 1605.00844v1

\bibitem{other} L. E. Ballentine : 
``The Statistical Interpretation of Quantum Mechanics'', {\em Rev. Mod. Phys.} {\bf 42} 385--380  (1970). 

\bibitem{o2} B. d'Espagnat : {\em Conceptual Foundations of Quantum Mechanics} $2^{nd}$ ed. (1976) Addison Wesley 

\bibitem{o3} D. Home and M. A. B. Whitaker : 
``The ensemble interpretation and context-dependence in quantum systems'',
{\em Phys. Lett. A} {\bf 115} 81--83  
(1986). 


\bibitem{EQM} C. L\'opez : ``An extended phase space for Quantum Me\-cha\-nics'', arXiv:1509.07025

\bibitem{dBro} L. de Broglie : 
``Interpretation of quantum mechanics
by the double solution theory'',
{\em Annales de la Fondation Louis de Broglie} {\bf 12}(4) 1--23 (1987)

\bibitem{subq} A. Khrennikov: ``A pre-quantum classical statistical model with
infinite-dimensional phase space'', {\em J. Phys. A: Math. Gen.} {\bf 38}, 9051
(2005).



\bibitem{s3} Gerard 't Hooft : {\em The Cellular Automaton Interpretation of Quantum Mechanics},
Fundamental Theories of Physics series, Volume 185 (2016), 
Springer Int. Pub. 





\bibitem{nogo} J. S. Bell :  
``On the Einstein Podolsky Rosen paradox'',
{\em Physics} {\bf 1}(3), 195--200 (1964). 

\bibitem{n2} S. Kochen and E.P. Specker: ``The problem of hidden va\-ria\-bles in quantum mechanics'',  {\em J.  Math.  Mech.} {\bf 17}, 59–87 (1967).
 
\bibitem{n3} J.F. Clauser, M.A. Horne, A. Shimony and R.A. Holt :
``Proposed Experiment to Test Local Hidden-Variable Theories'',
{\em Phys. Rev.  Lett.}{\bf 23}, 880--883 (1969).

 
\bibitem{n4} D. M. Greenberger, M. A. Horne, A. Shimony and A. Zeilinger : 
`` Bell's theorem without inequalities'',
{\em Am. J. Phys. } {\bf 58} (12) 1131--1142 (1990).

\bibitem{n5} N.D. Mermin : ``Simple unified form for the major no-hidden-variables theorems'', 
PRL {\bf 65}, 3373 (1990)






\end{thebibliography}
\end{document}